\documentstyle[12pt]{article}
\begin{document}
\baselineskip .8cm

\newcommand{\be}{\begin{equation}}
\newcommand{\ee}{\end{equation}}
\newcommand{\eps}{\varepsilon}
\newcommand{\z}{\zeta}
\newcommand{\cc}{\xi}

\title{Depinning transition of a directed 
polymer \\ by a periodic potential: 
a  $d$-dimensional solution.} 
\author{S. Galluccio and  R. Graber \\Institut de Physique 
Th\'eorique, Universit\'e de Fribourg,\\ CH -- 1700,  Switzerland}
\date{}
\maketitle

\abstract{ 
We study the depinning phase transition of  a directed
polymer in a $d$-dimensional space by a 
periodic potential localized on a  straight line.
We give exact formulas  in all dimensions for the critical pinning
we need  to localize the polymer.
We show that  a bounded state can still arise
even if, in average, the potential layer is not attractive
and for diverging  values of the potential on the repulsive sites.
The phase transition is of second order.
}

\vspace{1cm}
\newpage

The statistical mechanics of long linear 
chains (directed polymers) in ordered  and disordered 
media has been object of intense study in the past years. 
One  relevant problem in that large context is the study of  the
depinning transition of a directed  line
by a single extended defect embedded  in a  $(d+1)$-dimensional  lattice.
An attractive potential pins the interface (here a line) on itself
suppressing wandering, but  thermic fluctuations increase
the configurational entropy and a phase transition takes
place  at a given critical  temperature $T_c$ \cite{fln91}.
The depinning transition of polymers by a single 
defect has been object of intense work
and exact results are now well established in some simple
cases \cite{pff88},\cite{z2}. 
Moreover the simultaneous effect of both point and
extended defects has lead recently to some new results
in the context of RG approach in continuum models \cite{bk94}.

In this article we will deal with  the following problem: 
 the depinning phase transition of a single 
polymer in a $(d+1)$-dimensional  hypercubic lattice 
by a periodic potential localized on a line, i.e. a potential 
which is alternatively attractive and repulsive. 
We point out that such potential layer
can be used, for instance,  to mimic the effect of two 
alternating  kinds of pinning centers  with different strengths.
Moreover  it is tightly linked with a simplified  version 
of the KPZ equation for interface growth (see  discussion below).
Some well-known arguments show that with  
a $n$-dimensional (oriented) defect 
the polymer is localized for  $d-n+1 \le 2$ by an arbitrarily
weak attractive force, while for $d-n+1>2$ a finite strength
is necessary to do the work \cite{nv93}.
Analytical results  are  in general not available for 
such high-dimensional systems.

Our main results in this work are the following: 
(i) we solve the phase transition problem 
in {\it all} dimensions and we give an  
exact  formula for the critical   pinning strength   necessary
to localize the polymer at the origin.    
(ii) We surprisingly show that
a bound state can always arise even though, in average,
the potential layer is repulsive. This effect has been recently
 studied in the one-dimensional case \cite{nz95}.
We also prove that in {\it all}  
finite dimensions a finite strength on the attractive sites
is enough to pin the polymer even in the limit of an
infinite  potential  on the repulsive sites.
(iii) The  approach introduced is also
interesting by its own: we use a dual space 
representation of the transfer matrix which enables us to 
simply find  the critical state of the system and
the partition (wave) function.
This approach leads to exact results for  
the error-catastrophe problem in biological evolution \cite{ggz95}.

The energy of  a line of length $L$ with extremes at
 ${\bf h}^{(0)}=\vec{0}$ and ${\bf h}^{(L)}={\bf x}$, wandering 
in a $(d+1)$-dimensional space $\Omega={\bf Z}^d\otimes
{\bf N}$ and directed
along a  ``time'' axis, is given by \cite{fln91}:
\begin{equation}
{\cal H}(\{ {\bf h}^{(i)}\})= J \sum_{k=1}^L
\left|{\bf h}^{(k)}-{\bf h}^{(k-1)}\right|-\sum_{k=1}^L
U_k\delta_{{\bf h}^{(k)},\vec{0}} \quad ,
\end{equation}
where ${\bf h}^{(k)}$ is a vector  identifying the position
of the line in $\Omega$ at each ``time''  $k$. The potential is
localized at  the origin and it is alternatively attractive
and repulsive, i.e. $U_k=u>0$ if $k$ is even and $U_k=-v<0$ if
$k$ is odd. The directed line has no overhangs and RSOS 
condition is imposed.
A  canonical partition function is introduced (the sum
is over all possible allowed realizations of the interface
``height'' ${\bf h}^{(k)}$) :
\begin{equation}
{\cal Z}_L({\bf x})=\sum_{\{ {\bf h}\}}\exp\left(
-{\cal H}(\{ {\bf h}^{(k)}\})/T\right).
\end{equation}
 In the usual  approach one  defines a 
symmetrical transfer matrix ${\bf T}_{zz'}={\bf T}_{z-z'}$ from 
${\cal Z}$ as ${\cal H}=-T\ln \sum_z {\bf T}(z)\hat{S}_z$;
(here  $\hat{S}_z$ stands for a shift operator \cite{fln91}).
At finite temperatures $T>0$ the fluctuations of the interface
increase the configurational entropy while large humps
are unlikely since they give a higher internal energy.
The final state of the polymer is the result of that competition
and it is  associated 
to the free energy density (per unit length) $f$.
In the thermodynamic limit ($L \rightarrow \infty$) 
$f$ is dominated by the largest
eigenvalue of ${\bf T}_{zz'}$. 

For our system we see that in one step the
partition function ${\cal Z}_L({\bf x})$
obeys the following recursion relation:
\begin{eqnarray}
{\cal Z}_{L+1}({\bf x})&=&\left[1+(a_{L+1}-1)\delta_{{\bf x},\vec{0}}
\right]\\
& \times & 
\left[{\cal Z}_L({\bf x})+t\sum_{i=1}^L\left(
{\cal Z}_L({\bf x}+{\bf e}^{(i)})+
{\cal Z}_L({\bf x}-{\bf e}^{(i)}) \right) \right] \nonumber
\end{eqnarray}
where the unitary vectors ${\bf e}^{(i)}=(0,0,\cdots,i,\cdots,0)$
have a "1" bit as $i$-th. element  (so to satisfy RSOS conditions).
In the above we have defined the parameters $t=\exp(-J/T)\in (0,1]$
$a_L=\exp(u/T)$ (for $L$ even) and
$a_L=\exp(-v/T)$ (for $L$ odd).
We now introduce a dual space representation of our 
equation in order to simplify the calculation. For the present problem  
we  use a  standard Fourier transform, but 
sometimes one needs different representations \cite{ggz95}.
As the partition function ${\cal Z}({\bf x})$ is expected to 
be symmetric in the arguments
we introduce a cosine transform
\begin{equation}
{\cal Z}_L({\bf x}) =\int_0^1 {\rm d}^d k\, 
\prod_{i=1}^d\cos(\pi k_ix_i) {\cal Z}_L({\bf k}), \\
\end{equation}
and its inverse. In the Fourier space eq. (3) takes a simple form;
 a proper  definition of the transfer matrix should anyway take into
account the periodicity  of the problem.  If we introduce 
the  normalized quantities 
$G_L({\bf k})={\cal Z}_L({\bf k})/(1+2dt)^L $ and  
$\xi({\bf k})=(1+2t\sum_{i=1}^d \cos(\pi k_i))/(1+2dt)$,
after   two consecutive steps our  equation  reads:
\begin{eqnarray}
G_{2L+2}({\bf k}) & = & \xi^2({\bf k})G_{2L}({\bf k})+
A\int_0^1 {\rm d}^dq\; \xi^2({\bf q})G_{2L}({\bf q}) \nonumber\\
& + &  B\xi({\bf k})\int_0^1 {\rm d}^dq \; \xi({\bf q})G_{2L}({\bf q})
 \\
& + & 
 \frac{AB}{1+2dt}\int_0^1 {\rm d}^dq \;\xi({\bf q})\, G_{2L}({\bf q}).
\nonumber \end{eqnarray}
with  $A=a_{2L}-1$ and $B=a_{2L+1}-1$.

In order to find  the maximum eigenvalue $\eps$
we should consider the spectral equation
obtained from (5) by  identifying   
 $G_{L+2}({\bf k})=\eps^2G_L({\bf k})$.
 We recall that the only 
significant contribution
in the thermodynamic limit $L\rightarrow\infty$ is the
maximum eigenvalue  associated with (5).  In fact 
one can show  that $f\approx -\log \eps$  is always 
different from 0 in the localized region ($\eps>1$) while it 
vanishes in the 
unbounded  state for $\eps\rightarrow 1$. 
 The depinning phase transition
is  defined  at $\eps=1$ \cite{gg95}.   

The search for a general solution of the eigenvalue equation
is a very hard task,   nevertheless  one can   
find  the criticality condition as a function of the free
parameters $\{d,t,u,v\}$ of the  theory with no enormous effort. 
The idea is to   introduce two auxiliary  constants $K_{1,2}$
by integrating  all the terms containing the unknown function
$G_L({\bf k})$:
\be
K_n=\int_0^1 {\rm d}^d k\;  G_L({\bf k})\xi^n({\bf k})\quad (n=1,2).
\ee 
Therefore we get a homogeneous system  
of two algebraic  equations for $K_n$ 
which must  be satisfied by any general set of parameters $\{d,t,u,v\}$:
\begin{eqnarray}
\left(A{\cal I}_2(\eps)+\frac{AB}{1+2dt}{\cal I}_1(\eps)-1\right)K_1+
B{\cal I}_1(\eps)K_2 & = &0, \nonumber \\
\left(A{\cal I}_3(\eps)+\frac{AB}{1+2dt}{\cal I}_2(\eps)\right)K_1+
(B{\cal I}_2(\eps)-1)K_2&=&0,  
\end{eqnarray}
with
\be
{\cal I}_n(\eps)=\int_0^1 {\rm d}^dk\;  \frac{\xi^n({\bf k})}
{\eps^2-\xi^2({\bf k})} \quad (n=1,2,3).
\ee
The homogeneous system (7) admits  non trivial solutions
iff the determinant of its coefficients vanishes. This is therefore
the condition we must require in order to get the spectrum
of the transfer matrix. 
Performing the limit $\eps\rightarrow 1$
we arrive at the condition which must be satisfied  by any 
set  $\{d,t,u,v\}$
at the critical point \cite{gg95}. After some 
 calculations we then get the criticality condition:  
\be
1-(A+B){\cal I}_2'+AB\left[{\cal I}_2'^2-{\cal I}_1'\left({\cal I}_3'
+\frac{1}{1+2dt}\right)\right]=0,
\ee
with ${\cal I}_n'={\cal I}_n(\eps=1), \quad (n=1,2,3)$. In the following 
we will drop  the ``prime'' from the formulas, anyway recalling
that all quantities are calculated at $\eps=1$.
Borrowing from the thermodynamic
language, equation  (9) can be thought as the 
{\it equation of state} at criticality: the  ``thermodynamic
variables'' are now those  in the set $\{d,t,u,v\}$.

Despite of the complexity of the high dimensional 
integrals ${\cal I}_n$ involved in the above formula 
one can finally express them,  after  some analytical work,
in the following form:
\be
{\cal I}_1=\frac{\alpha}{2}(f-g),\quad {\cal I}_2=\frac{\alpha}{2}(f+g)-1,
\quad {\cal I}_3={\cal I}_1-\frac{1}{2t\alpha},
\ee
where we have defined a new constant $\alpha=(1+2dt)/2t$ and
the two integrals:
\begin{equation}
f  = \int_0^1 {\rm d}^d k\; \frac{1}{d-\sum'({\bf k})}, \quad
g  = \int_0^1 {\rm d}^d k\; \frac{t}{1+dt+t\sum'({\bf k})}.
\end{equation}
with $\sum'=\sum_{i=1}^d\cos(\pi k_i)$.
By means of the above definitions our equation of state
reads 
\be
A=\frac{B{\cal I}_2-1}{B({\cal I}_2^2-{\cal I}_1^2)-{\cal I}_2}.
\ee
The integrals $f$ and $g$ are well known 
in the theory of random walks (RW)
\cite{yd}; the former, in particular,  has a well defined physical
meaning: it gives the {\it mean time} spent on the origin for
a random walker  in a $d$-dimensional hypercubic lattice
(times $1/d$).  In  other words $1-d/f$ is the probability 
of return of a random walker to his starting point. 
We use an  integral representation  to write them  in a simpler
form:
\be
f=\int_0^{\infty}{\rm d}u\; e^{-du} I_0(u)^d,\quad
g=\int_0^{\infty}{\rm d}u\; e^{-\left(d+1/t\right)u} I_0(u)^d.
\ee
Here $I_0(u)$ is the usual modified Bessel function
of integer order.
The mathematical properties of $f$ and $g$ are central to
our solution and then we will summarize them in more detail.
 All below results hold in the ranges: 
$d\in [0,\infty)$ and $t\in (0,1]$ \cite{gg95}.

Both $f$ and $g$ are positive strictly convex decreasing functions 
of $d$, converging to $0^+$
for $d\rightarrow\infty$. Moreover we have that  $f >g$
$\forall t\in (0,1]$.  It is interesting
to look at their behavior in some 
extreme situations.  If we perform an  asymptotic development
of $g$ for $t$ close to 0 we get the result $g= t-dt^2+O(t^3)$.
Moreover we find that
\be
f=\frac{1}{d}\left(1+\frac{1}{2d}+\frac{3}{4d^2}+\frac{3}{2d^3}
+O\left(\frac{1}{d^4}\right)\right)
\ee
\be
g=\frac{1}{d+1/t}+\frac{d}{2(d+1/t)^3}+\frac{3d^2}{4(d+1/t)^5}
+O\left(\frac{1}{d^4}\right)
\ee
from the  asymptotic developments at large $d$.  
Perhaps the most important  property  of the two above integrals
 is that $f$ is a divergent integral for $d=1,2$ 
 while it is finite for $d>2$ (e.g. see \cite{yd}).   
The second one, $g$, is  finite, on the other hand, $\forall d$.
 In the RW theory the divergence of $f$ for $d<3$ leads
to the well-known  result that  the total probability
or return of a random walker to his starting point 
is 1 only for dimensions
less than 3. Our solution   shows directly  how this
pure topological effect
 plays a central role in the context  of  the depinning
transition for directed polymers.

For $d=1$,  $g$ can be explicitly calculated and  by taking
the dominant contribution of (12) at diverging $f$,
the criticality condition becomes
\be
\frac{u}{T}=\log\left(1-\frac{B}{1+B\left(2-\sqrt{1+2t}\right)}\right)\, ,
\ee 
which confirms the result obtained by Nechaev
and Zhang in the same context \cite{nz95}.
Let us now turn back to eq. (12) for the general case. At $t=0$,
or equivalently $J\rightarrow \infty$, the polymer is a rigid
straight line and then it is in the pinned (resp. unpinned) phase 
depending on the sign of the difference $u-v$. 
This result  is indeed contained  in our solution: 
by asymptotically expanding $g$ for small $t$ (see  above)
we finally find
\be
A=-\frac{B}{B+1}+\frac{4+4B-B^2(df-1)}{f(B+1)^2}t+O(t^2).
\ee
We see that, as expected, at vanishing hopping constant $t$ 
there is no more dependence on the dimensionality;  
and solving for the pinning 
strength one gets the result that $A(B+1)=-B$ or
 $u=v$. The ``phase diagram'' on the plane $u$-$v$ is then
represented in this case by a single straight line bisecting the 
whole space (see Fig.(1)).
In the above semispace ($u>v$) the polymer is in a bound
state, while in the lower one ($u<v$)  it is 
completely delocalized.

What does happen if the hopping constant $t$ is different from 0?
In, by now, standard notation we define $u_c$ as the value
of the force on the attractive sites which satisfies (12)
for a given set $\{d,t,v\}$.
For $B=0$, or equivalently $v=0$, the potential layer is made
of alternating attractive  and neutral sites and the criticality
condition reads $u_c=T\log(1+1/{\cal I}_2)$.  This is an exact
formula  valid in all dimensions.    
Since $f$ (and then ${\cal I}_2$) diverges for $d<3$, 
an arbitrarily small $u=\delta>0$ is enough 
to localize the polymer, according
to well-know general results, while for $d \ge 3$ 
we  need  a  finite value
\be
e^{u_c/T}=2d+\frac{-1+2t-2t^2}{t^2}
+\frac{1-2t+t^3-3t^4}{2t^4}\frac{1}{d} +O\left[\frac{1}{t^6d^2}\right],
\ee
to do the job (this  phenomenon  has  also a QM counterpart
 \cite{ll}). 
Moreover that critical attractive force diverges  
logaritmically at large $d$.
The same result we  get in the 
more familiar case $v=-u$ representing an extended linear 
attractive defect. In this case one finds
\be 
\frac{u_c}{T}=\log\left[1+\frac{1}{{\cal I}_1+{\cal I}_2}\right],
\ee
 which gives,
at $d=1$, a known result, i.e. $u_c=0$. Fig.(2) shows the shape
of $u_c/T$ as a function of the dimension $d$ in that case.      

An interesting aspect of our solution is that a bound
state can take place at {\it all} finite dimensions $d$ 
even if  the potential layer is, in average, repulsive ($v>u$).
This is evident from Fig.(1) in which we have drawn 
the critical curves separating bounded and unbounded
regions for different values of $d$ as a function of the 
``reduced parameter''  $v/T$. 
Above (below) the straight line $u=v$ the potential is, in average,
attractive  (repulsive).  Then we see that for every fixed
$v$ and $\forall d$ (finite), one can find 
a finite value of $u_c<v$ which localize the polymer giving
a bound state for the partition function. 
The more astonishing point is that one can simply prove from (12) 
that the critical curves asymptotically converge toward a finite
$u_c$ also in the extreme  limit $v\rightarrow +\infty$:
\be
\frac{u_c^{(\infty)}}{T}= \log\left[1+
\frac{1+{\cal I}_2}{({\cal I}_2^2-{\cal I}_1^2)+{\cal I}_2}
\right]\, .  
\ee
Again, for $d=1$, we recover the known result \cite{nz95}
$\exp(u_c^{(\infty)}/T)=\sqrt{1+2t}/(\sqrt{1+2t}-1)$.
The intuitive explication of this apparent paradox is that
the polymer wanders in the space avoiding repulsive sites and 
passing through the potential layer  on the attractive
ones (preferentially). 
Some critical curves are showed in Fig.(1)  for both $d<2$
and $d>2$ as functions of the reduced parameter $v/T$.

We recall that, as obvious, instead of choosing  $u$ and $v$ as ``free
parameters'' in the above considerations, we could, 
in principle,   directly look at the behavior of the system
as a function of the temperature $T$ (from eq.(12)) with
$u$ and $v$ fixed.

In presence
of a potential layer the  broken translational symmetry 
of $\Omega$ can be associated to the presence of
Goldstone modes whose mass $\mu$ is finite for $T<T_c$
(i.e. in the pinned phase).  The 
bounded state can be expressed as
${\cal Z}({\bf x})=\exp(-\mu|{\bf x}|)$ and the maximum
eigenvalue as $\eps= 1+\mu^2$.
By using the explicit form of the partition
(wave) function in the localized phase \cite{gg95}
  we find that near the transition $|f-f_c|$ vanishes
quadratically with $(u-u_c)$, i.e. the transition is of 
second order.
Then the transversal correlation length $\xi_{\perp}$ 
diverges as $(u-u_c)^{-1}$ at the critical point.

Conclusion:  we have studied  the problem of a polymer in a 
$d$-dimensional space  in presence of a 
linear extended defect with periodically arranged
pinning sites and we have found the exact  condition
for  occurring of the depinning transition.
Exact formulas for the critical pinning are found in all
dimensions.
Perhaps the most interesting aspect of our system is that
a bounded state of the partition function can arise  for all $d$
even if the potential on the extended defect
is  in average repulsive.
We  could  also get  a deeper insight
into the physical problem from  the calculation of the
of  the partition wave function. 
This seems to be a very interesting aspect which can also
be dealt with our approach \cite{gg95}.
Recently has been pointed out  that the 
physics of  a  polymer subjected to a 
linear extended defect  can be seen, in the context 
of  dynamical growing of interfaces, 
as a simple version of the 
KPZ equation \cite{kpz} with a ``noise'' term 
proportional to 
$\delta^d({\bf x})$ \cite{nk}.   
 We will  present  our   results on these problems
 in a separate paper.

We would like to thank Prof. Yi-Cheng Zhang for useful
comments and suggestions. This work was supported
by the Swiss National Fund for Scientific Research.

\newpage

\newpage
\section*{Figure captions}

{\bf Fig. 1}\newline
Critical curves $u_c/T$ vs. $v/T$ for different
values of $t$ and $d$ calculated by numerically integrating
$f$ and $g$ (see text).
Above (below) them the system 
is in a bounded (unbounded) state for the
partition function. At vanishing hopping constant ($t$=0)
the depinning line is  the diagonal of the phase space $u$-$v$.
For $t >0$ and $d>2$ we need a finite  value of
the potential $u$ to pin the polymer.
\vskip .7cm
{\bf Fig. 2}\newline
The critical pinning (divided by $T$) necessary to localize
the polymer as a function of $d$ in the 
case of a uniform attractive potential (i.e. $v=-u$).
 The divergence of $u_c/T$ with $d$ is logarithmic (see text).
\vskip .7cm 

\end{document}